\documentclass[utf8]{frontiersSCNS} % for Science, Engineering and Humanities and Social Sciences articles
\usepackage{url,hyperref,lineno,microtype,subcaption}
\usepackage[onehalfspacing]{setspace}
\usepackage{comment} % To enable section comments
\usepackage{svg} % To enable svg images. But in the submitted version images have to be of a different type!
\usepackage{graphicx}
\usepackage{float}

\usepackage{todonotes}

\def\keyFont{\fontsize{8}{11}\helveticabold }
\def\firstAuthorLast{Kronborg {et~al.}} %use et al only if is more than 1 author
\def\Authors{Joel Kronborg\,$^{1,*}$, Frida Svelander\,$^{1}$, Samuel Eriksson-Lidbrink\,$^{1}$, Ludvig Lindström\,$^{1}$, Carme Homs-Pons\,$^{1}$, Didier Lucor\,$^{2}$ and Johan Hoffman\,$^{1}$}
\def\Address{$^{1}$KTH Royal Institute of Technology, Department of Computational Science and Technology, School of Electrical Engineering and Computer Science, Stockholm, Sweden 
\\ $^{2}$Laboratoire Interdisciplinaire
des Sciences du Num{\'e}rique, LISN-CNRS, Orsay, France}
\def\corrAuthor{Joel Kronborg}

\def\corrEmail{joelkro@kth.se}

\begin{document}
\onecolumn
\firstpage{1}

\title[Computational turbulent ventricular blood flow]{Computational analysis of flow structures in turbulent ventricular blood flow associated with mitral valve intervention} 

\author[\firstAuthorLast ]{\Authors} %This field will be automatically populated
\address{} %This field will be automatically populated
\correspondance{} %This field will be automatically populated

\extraAuth{}% If there are more than 1 corresponding author, comment this line and uncomment the next one.
%\extraAuth{corresponding Author2 \\ Laboratory X2, Institute X2, Department X2, Organization X2, Street X2, City X2 , State XX2 (only USA, Canada and Australia), Zip Code2, X2 Country X2, email2@uni2.edu}

\maketitle

\begin{abstract}

%%% Leave the Abstract empty if your article does not require one, please see the Summary Table for full details.
%\section{}

Cardiac disease and clinical intervention may both lead to an increased risk for thrombosis events due to a modified blood flow in the heart, and thereby a change in the mechanical stimuli of blood cells passing through the chambers of the heart. Specifically, the degree of platelet activation is influenced by the level and type of mechanical stresses in the blood flow. 
In this article we analyze the blood flow in the left ventricle of the heart through a computational model constructed from patient-specific data. The blood flow in the ventricle is modeled by the Navier-Stokes equations, and the flow through the mitral valve by a parameterized model which represents the projected opening of the valve. A finite element method is used to solve the equations, from which a simulation of the velocity and pressure of the blood flow is constructed. The intraventricular blood flow is complex, in particular in diastole when the inflow jet from the atrium breaks down into turbulent flow on a range of scales. 
A triple decomposition of the velocity gradient tensor is then used to distinguish between rigid body rotational flow, irrotational straining flow, and shear flow. The triple decomposition enables the separation of three fundamentally different flow structures, that each generates a distinct type of mechanical stimulus on the blood cells in the flow. 
We compare the results to a simulation where a mitral valve clip intervention is modelled, which leads to a significant modification of the ventricular flow. Further, we perform a sensitivity study of the results with respect to the positioning of the clip. 
It was found that the shear in the simulation cases treated with clips increased more compared to the untreated case than the rotation and strain did. A decrease in valve opening area of 64~\% in one of the cases led to a 90~\% increase in rotation and strain, but a 150~\% increase in shear.
The computational analysis suggests a process for patient-specific simulation of clinical interventions in the heart with a detailed analysis of the resulting blood flow, which could support clinical risk assessment with respect to platelet activation and thrombosis events. 

\tiny
 \keyFont{ \section{Keywords:} patient-specific heart modelling, left ventricle, mitral valve clip, finite element method, turbulent blood flow, triple decomposition of velocity gradient tensor} %All article types: you may provide up to 8 keywords; at least 5 are mandatory.
\end{abstract}

\section{Introduction}
%For Original Research Articles \citep{conference}, Clinical Trial Articles \citep{article}, and Technology Reports \citep{patent}, the introduction should be succinct, with no subheadings \citep{book}. For Case Reports the Introduction should include symptoms at presentation \citep{chapter}, physical exams and lab results \citep{dataset}.

Heart attack and stroke follow from a loss of blood supply to the heart and brain, respectively, often caused by a blood clot that obstructs a critical blood vessel, referred to as thrombosis. Blood clots are naturally formed as a response to injury to avoid blood loss, but may also form under other circumstances where the platelets (thrombocytes) of the blood are activated which leads to an increased tendency of the platelets to aggregate, see e.g., \cite{previtali2011risk}. 
%, hence,  an elevated risk of blood clot formation. 
%
Platelet activation can be triggered by contact with a foreign material, such as a medical device, see e.g., \cite{rojano2019kinetics}, or by the mechanical stresses in the blood flow. 
Specifically, it is associated with elevated levels of shear in the blood. The degree and time extent of the activation is a function of the type of mechanical stimulus experienced by the platelets, see e.g., \cite{rahman2020effects}. Therefore, a detailed analysis of the structure of the blood flow may help to assess the risk for platelet activation and thrombosis. 
The heart is supplied blood through the right and left coronary arteries, with inlets through the right and left coronary sinuses which are part of the aortic root attached to the left ventricle (LV) of the heart. Further downstream the aorta, the brain is supplied blood through the carotid arteries that emanate from the aortic arch. As a consequence, the level of platelet activation in the LV blood flow is of interest when assessing the risk for thrombosis events in both the heart and the brain. 

Blood flow in the LV is complex, in particular in diastole when the heart relaxes and blood flows into the ventricle from the left atrium through the open mitral valve. The ventricular blood flow is also highly sensitive to the structure and function of the mitral valve, and pathologies such as mitral valve regurgitation or stenosis can drastically change the flow pattern. In the early filling phase of diastole (the E-wave) a jet is formed which then breaks down in the diastasis phase, after which a second jet is formed in the atrial contraction phase (the A-wave). The two successive jets create shear layers and vortex rings in the ventricle, which then break down into turbulent flow structures on a range of scales. 
The blood flow generates mechanical stimuli on blood cells that trigger different responses in the cells, for example, activation of platelets. 
Mechanical stimuli on blood cells are difficult or impossible to assess by medical imaging alone, and the turbulent nature of the ventricular flow makes this challenge even greater. In this article we investigate to what degree a computational model can be used to assess the mechanical stimuli on blood cells in the LV flow, building on our previous work on patient specific simulations of the LV blood flow, see e.g., \cite{larsson2017patient} and \cite{larsson2017estimation}, and analysis of turbulent flow structures, see \cite{hoffman2021energy}. 

Cardiac disease and clinical interventions may both lead to changes in the LV blood flow and the mechanical stimuli experienced by the blood cells in the ventricle. 
Mitral valve regurgitation is a condition which affects the LV blood flow, where the mitral valve does not close properly and therefore may leak in systole when the valve is supposed to be closed. A common procedure to mitigate mitral valve regurgitation is to suture the two leaflets of the mitral valve together with a a mitral valve clip (MVC).
While a MVC intervention reduces or removes the regurgitation, there is a risk for creating a new problem in the form of stenosis, due to a reduced valve opening. 
Recently, patient-specific computer simulations have been used to evaluate the risks and benefits of a MVC intervention, see \cite{caballero2020comprehensive}.  
In this article we use a finite element method to simulate the blood flow in the LV before and after a MVC intervention, to analyze the difference in the ventricular blood flow. We also perform a sensitivity study with respect to the positioning of the MVC. 
A two dimensional model of the mitral valve is used, which represents an approximation of the projected opening between the left atrium and the left ventricle. The mitral valve model is then modified to take the positioning of the MVC into account. 

To analyze the simulated blood flow we use the triple decomposition of the velocity gradient tensor, suggested by  \cite{kolavr2007vortex} to improve the identification and visualization of vortices. 
\cite{hoffman2021energy} noted that the triple decomposition implies that the flow locally can be approximated by a sum of shear flow, rigid body rotational flow and irrotational straining flow. 
Our hypothesis is that this refined analysis of the blood flow will lead to an improved quantification of the type and extent of the mechanical stresses experienced by blood cells while inside the LV, which can support a risk assessment of thrombosis events to a greater extent than a blood flow analysis based solely on the velocity and pressure of the blood. 

In Section~\ref{section:method} we describe the computational model and the triple decomposition of the velocity gradient tensor, and in Section~\ref{section:results} we present results from the simulations. These results are then discussed in Section~\ref{section:discussion}.

% More extensive description of mitral regurgitation and the mechanical effects on blood?

\section{Method}
\label{section:method}

\subsection{Left ventricle model}
% Short version of section II.A in Larsson et al?
% In/outflow governed by the change in volume
The LV model is based on 4D transthoracic echocardiography (TTE) images of the left ventricle of one human subject, acquired using the technique described in detail in \cite{larsson2017patient}. Images were captured in 26 frames over a full cardiac cycle which lasted for 830 ms, and the heart wall was segmented in a semi-automated fashion in each image, see \cite{hansegaard2009semi}. From this a set of triangulated surface meshed were generated which described the deformation of the endocardium over the cardiac cycle. 

The orifice of the mitral valve (MV) was identified by a sonographer in thirteen 2D short axis B-mode views in early diastole, with each view rotated 15 degrees around the LV long axis. Similarly, the aortic valve opening was identified in a single 3-chamber B-mode view in early systole. Both valve openings were marked on the surface meshes.

From one surface mesh corresponding to end diastole, here referred to as the initial surface mesh, a refined version was produced, from which a tetrahedral volume mesh of the interior of the ventricle was generated. This volume mesh consists of 406,964 vertices. This resolution was chosen based on the  convergence studies in \cite{larsson2017patient}, which showed less than 2~\% deviation in results when increasing mesh resolution above 400,000 vertices.

\subsection{Simulation of blood flow}
% Incompressible, Newtonian
The blood in the LV is modelled as an incompressible Newtonian fluid. Although blood is known to express non-Newtonian properties in smaller blood vessels, see e.g., \cite{pedley1980fluid}, and \cite{adjoua2019blood}, the blood inside the heart chambers is typically modelled as a Newtonian fluid, see \cite{westerhof2010snapshots}.
To simulate the LV blood flow, we determine the velocity vector $\mathbf{u}(\mathbf{x},t) : \Omega^t \rightarrow \mathbb{R}^3$ and scalar pressure $p(\mathbf{x},t) : \Omega^t \rightarrow \mathbb{R}$, such that the Navier-Stokes equations are satisfied: 
\begin{eqnarray}
&\rho (\dot{\mathbf{u}} + (\mathbf{u} \cdot \nabla)\mathbf{u}) - \mu\Delta\mathbf{u} + \nabla p = 0, &(\mathbf{x}, t) \in \Omega^t \times [0,T] \label{nse1}\\
&\nabla \cdot \mathbf{u} = 0, &(\mathbf{x}, t) \in \Omega^t \times [0,T],\label{nse2}
\end{eqnarray}
defined over the time-dependent domain 
$\Omega^t\subset \mathbb{R}^3$ at time $t\in [0,T]$. 
Equation (\ref{nse1}) expresses conservation of momentum, and equation (\ref{nse2}) conservation of mass. Standard mathematical notation is used, with $\nabla $ the gradient operator and $\Delta $ the Laplacian operator. We use the dynamic viscosity $\mu = 0.0027$ Pa$\cdot$s and the blood density $\rho = 1060$ kg/m$^3$, see  \cite{di2001fluid}.

$\Omega^t$ represents the deforming LV, with the boundary $\partial \Omega ^t$ partitioned into the heart wall and the two valve orifices. 
We use no slip velocity boundary conditions, corresponding to setting the blood flow velocity equal to the deformation velocity of the heart wall and closed valves. In systole, the open aortic valve (AV) is modelled by a standard outflow boundary condition, where the pressure is set to zero. 
The model of the mitral valve is described below, with and without a MVC intervention. 
To simulate the LV blood flow over the cardiac cycle, the domain $\Omega^t$ is approximated by the tetrahedral volume mesh which is deformed over time based on the 26 surface meshes. For a detailed description of this process, see \cite{larsson2017patient}. The Navier-Stokes equations are solved on this deforming volume mesh using an aribtrary Lagrangian-Eulerian (ALE) formulation of a stabilized finite element method, which is described in \citet{hiromi2021interface}. 

\subsection{Mitral valve model}

We extend the LV model with a simple planar, data-driven parameterized model of the MV dynamics, which acts as a time dependent inflow boundary condition.
%in the Heartsolver framework. 
The MV is modeled as a time-varying orifice representing the area enclosed by the planar projection of the leaflets onto the mitral annulus, accounting for virtual folding/unfolding mitral leaflets, similar to the work in \cite{imanparast2016impact}.

\begin{figure}[h!]
\begin{center}
\includegraphics[width=10 cm]{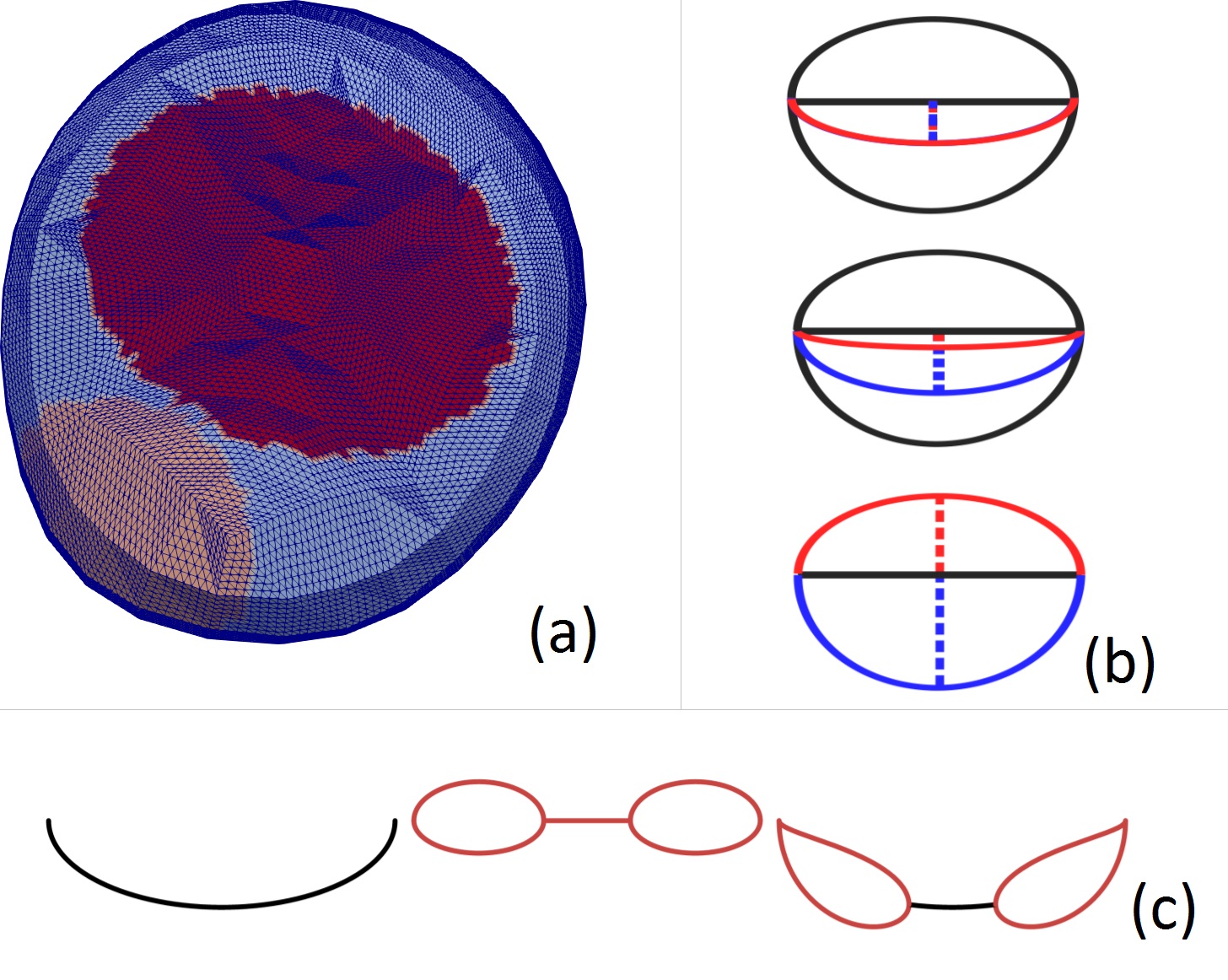}
\end{center}
\caption{(a) AV placement (yellow) and MV placement (red) on the LV surface mesh in an axial view. (b) The two half ellipses making up the MV model at different stages throughout the heartbeat cycle. The solid red line is the projection of the anterior leaflet, and the solid blue line the posterior. The dashed lines are the short axes of the corresponding half ellipse. In the top view the valve is fully closed (systole), and the leaflets overlap on the coaptation line. In the bottom view the valve is fully open (E-wave peak), and the center view shows early diastole. The solid black line through the valve model is the long axes of the two half ellipses. (c) The coaptation line (left), the two double half ellipses making up the MV model after MVC intervention (center) and the final shape of the MV model after MVC intervention (right). The part that does not open is the location of the clip.}
\label{fig:MV_mesh}
\end{figure}

The basic shape of the annulus is modeled as two half ellipses sharing a common long axis. From the ultrasound identification of the MV, the length of the long axis and the two short axes are identified, as well as the spatial 2D location of the coaptation line within the annulus area. An MV center point on the wall of the ventricle mesh is also given from the TTE images, which is used as a spatial reference point on the mesh to keep the MV in place. In figure \ref{fig:MV_mesh}a the MV and AV areas on the simulation mesh are displayed, as well as the contour of the MV model and the projected leaflets for a particular moment in time in figure \ref{fig:MV_mesh}b.

The opening and closing of the MV are given as temporal variations of the short axes of the two half ellipses. During systole, the short axes are equal, effectively keeping the projection of the leaflet edges overlapped along the coaptation line. Throughout diastole, the short axes are scaled to imitate a projection of the moving leaflets onto the surface mesh. Effectively this corresponds to a maximum opening area at E-wave peak, slightly smaller in diastasis, and somewhere in between at A-wave peak. For these points in time in the diastole timeline, the short axis scalings are determined by an idealized MV, see \cite{fyrenius2001major}. Patient-specific scalings are difficult to determine from echocardiography images since the resolution is limited. The predetermined scale factors are given in table \ref{table:diastole_timings} for E-wave peak, diastasis start and end, and A-wave peak.

\begin{table}[h!]
\begin{center}
\caption{Timings of different parts of diastole, and scalings of the short axis anterior (SAA) and short axis posterior (SAP) of the MV model at those times. The timings are determined by analyzing changes in the rate of expansion of the LV. Note that the coaptation line does not coincide with the intercommissural line (long axis), but the scalings at diastole start and end correspond to both leaflet edges coinciding with the coaptation line, i.e. the MV is closed at these times.}
\begin{tabular}{ |l|c|c|c|c|c|c| } 
 \hline
  & Diastole start & E-wave peak & Diastasis start & Diastasis end & A-wave peak & Diastole end \\
 \hline
 $t$ (s) & 0.420 & 0.525 & 0.644 & 0.711 & 0.779 & 0.830 \\
 \hline 
 SAA scale & 0.40 & 1.00 & 0.50 & 0.50 & 0.85 & 0.40 \\
 \hline
 SAP scale & -0.57 & 1.00 & 0.70 & 0.70 & 0.85 & -0.57 \\
 \hline
\end{tabular}
\label{table:diastole_timings}
\end{center}
\end{table}

Although the patient-specific scaling of the short axes could not be determined from the echocardiography images, the timings for the E-wave and A-wave peaks, as well as diastasis start and end, are determined based on the expansion rate of the LV of the particular subject studied here. This can be done patient specifically for other cases as well.

\begin{figure}[h!]
\begin{center}
\includegraphics[width=\linewidth]{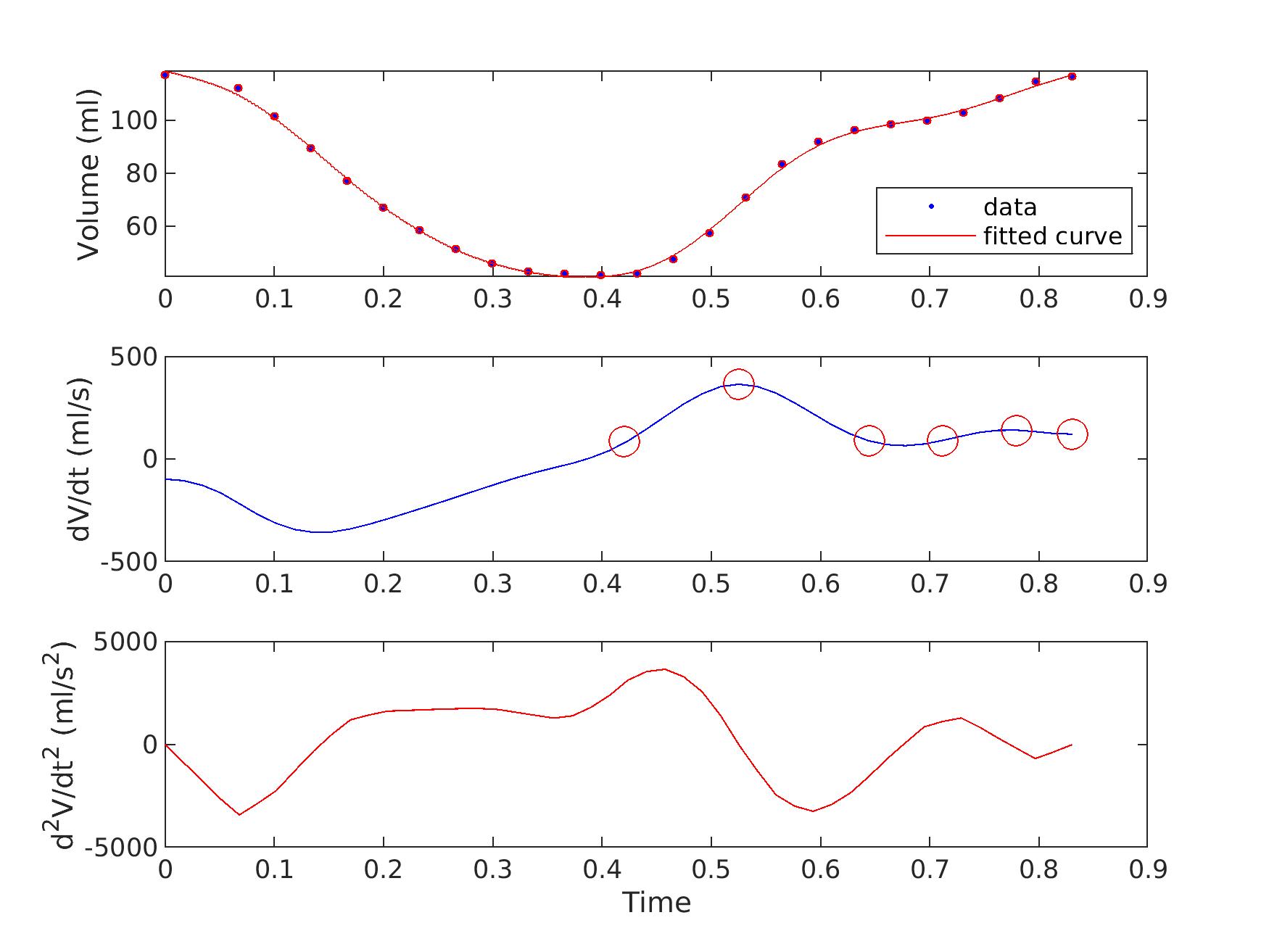}
\end{center}
\caption{Volume of the acquired surface meshes of the LV, and the first and second derivative of the interpolated curve. The top plot shows the volumes of the 25 used frames directly from the TEE images as blue dots, and the interpolation as a red line. The middle plot shows the first derivative, and the resulting timings of diastole start, E-wave peak, diastasis start and end, A-wave peak, and diastole end as red circles (see table \ref{table:diastole_timings} for exact values). The bottom plot shows the second derivative.}
\label{fig:volume_vs_time}
\end{figure}

The volumes of the 26 frames extracted from the echocardiography images are displayed in figure \ref{fig:volume_vs_time}, plotted against time from the beginning of systole. A cubic smoothing spline curve, $V(t)$ is fitted to the data points. From the first and second derivative (also included in figure \ref{fig:volume_vs_time}) of this smooth curve the timings are identified. Fundamentally, the shift from systole to diastole is identified as the point where the first derivative $dV/dt$ (i.e. the rate of the volume change) is 0, shifting from negative (contraction) to positive (expansion). The shift from diastole to systole is given simply as the beginning/end time of this heartbeat. Note that the isovolumetric phases between diastole and systole are excluded for simplicity.

% Fix the y labels

The E-wave and A-wave peaks are easily identified as the local maxima in $dV/dt$, with E-wave peak being the former and higher one. To define diastasis, the local minimum between these two peaks is determined. Diastasis is then defined as the time between the two original data frames from the TTE images on either side of this local minimum. These timings are listed in table \ref{table:diastole_timings}.

\begin{figure}[h!]
\begin{center}
\includegraphics[width=\linewidth]{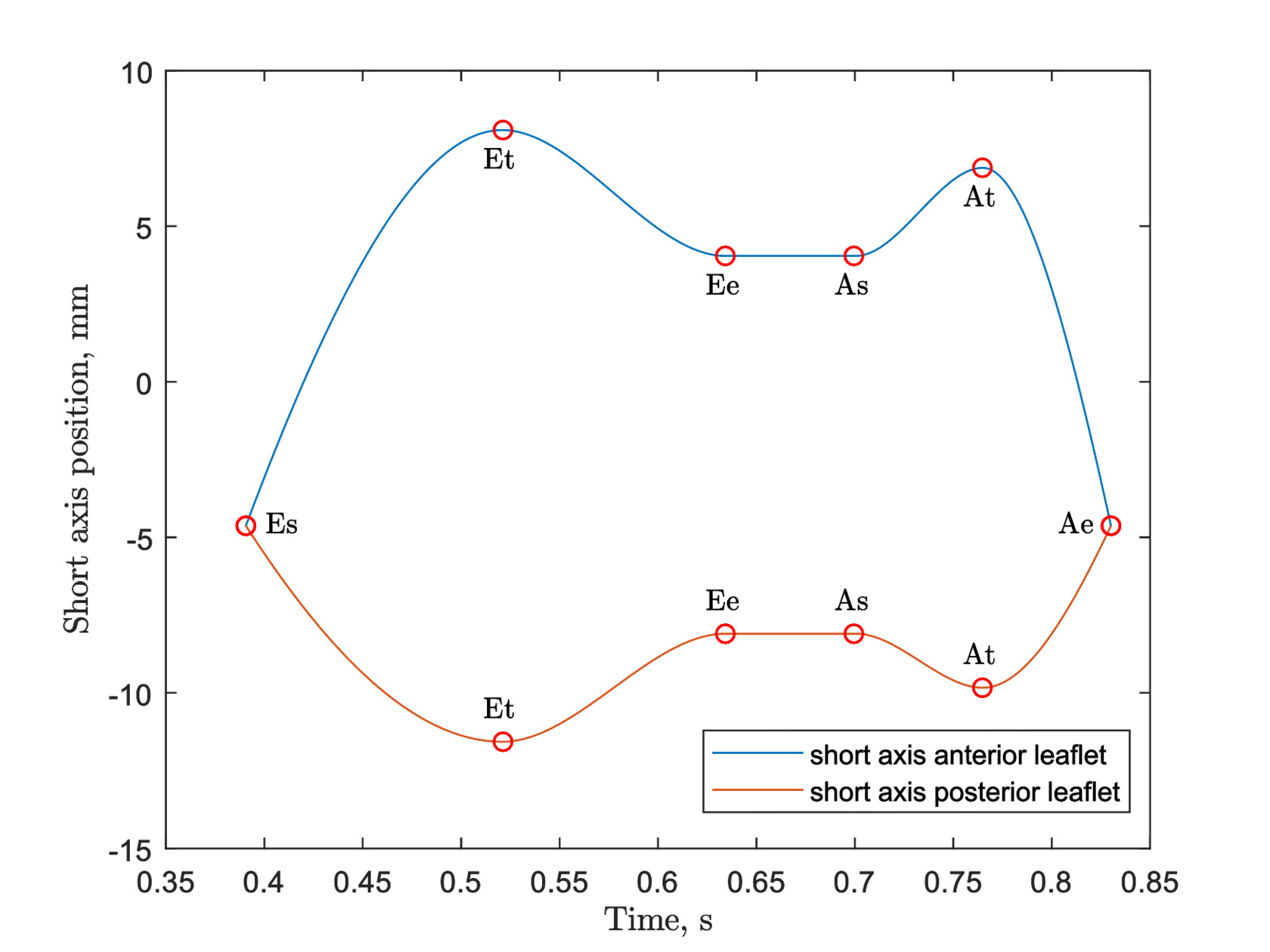}
\end{center}
\caption{Lengths of the short axes of the anterior (blue) and posterior (red) leaflets in the MV model over time. The times marked by red circles are E-wave start (Es), E-wave peak (Et), E-wave end/diastasis start (Ee), A-wave start/diastasis end (As), A-wave peak (At) and A-wave end (Ae). The times and scalings of the short axes are given in table \ref{table:diastole_timings}.}
\label{fig:mvtimings}
\end{figure}

With the timings of these four crucial points determined, along with the timings for shifts from diastole to systole and vice versa, the scaling of the short axes of the MV model are given by cubic Hermite interpolation between the scaled axis lengths and times given in table \ref{table:diastole_timings}. The resulting axes scaling over the full duration of diastole is displayed in figure \ref{fig:mvtimings}.

% Timings, including figure (and table?). This is straight from Frida's draft, needs to be expanded
%The opening and closing pattern for the MV is based on timings extracted from the variation in volume in the surface mesh data. Six points in time in diastole were identified: opening, E-wave peak, diastasis start and end, A-wave peak, and closing. These are marked in figure \ref{fig:mvtimings}. Opening and closing, i.e. diastole start and end, were identified as the time when expansion of the chamber started and ended respectively. The point at which the rate of expansion $dV/dt$ was greatest was identified as the E-wave peak. Diastasis start and end times were based on the slope of the expansion rate curve, and A-wave peak was identified as the point of most rapid expansion between diastasis end and MV closing.

% size of opening in each "state". Ask Didier for sources? Seems like there are some in box

\subsubsection{After mitral valve clip intervention}

To simulate the inflow of blood after a MVC intervention, part of the coaptation line of the MV was kept stationary in its closed state throughout diastole. This produces a valve model with two smaller openings, one on either side of the clip as displayed in figure \ref{fig:MV_mesh}c. This is achieved by creating two smaller openings consisting of two half ellipses each, similar to the untreated case, and superpositioning this with the coaptation line. The two smaller openings behave much like smaller versions of the untreated MV model throughout diastole. They follow the same pattern for opening and closing, i.e. the same scalings and timings given in table \ref{table:diastole_timings}. In its closed state, the clipped MV model has the same coaptation line as the untreated valve model.

Since the length of the coaptation line is the same for the clipped and untreated models, the long axes of the two smaller openings are directly given from the remaining unclipped parts. Additionally, the relationship between the anterior and posterior short axis of each opening is the same as the relationship of the short axes in the untreated case. The length of the two anterior short axes are assumed to have the same relationship as the length of their corresponding long axes, and the same holds for the posterior short axes.

To compute the maximum length of the short axes in the MVC case, the length of the edge of the openings at E-wave peak is assumed to be the same as the length of the edges in the untreated case, i.e. the circumference of the orifice. Along with the relationships between different axes described above, this is used to uniquely determine the maximum length of the short axes. As in the untreated case, this is assumed to imitate a projection of the 3D MV opening onto the surface mesh of the ventricle, see \cite{imanparast2016impact}. Given the length of the ellipse axes and the MVC, the total area of the two openings were computed and are presented in table \ref{table:clip_pos}.
%Following the standard width of the MVC NT device\citep{mitraclip_description}, the width of this static portion of the coaptation line was set to 5 mm. On either side of the clip the MV model used the same opening timings as for the untreated valve.
% Figure of the opening profile? Figure 10 from kex

\subsection{Simulations}
% left = anterolateral, right = posteromedial
Five different cases were simulated, with MVC centers positioned along the coaptation line at $s\cdot l$ from the coaptation line center, with $l$ being the same long axis length as used in the untreated valve, i.e. half the intercommissural distance, and $s\in (-1, 1)$ a scaling parameter. The values of $s$ used in these simulations are given in table \ref{table:clip_pos}, along with the resulting MV opening areas at mid diastole. Hence, $s = 0$ corresponds to the clip being in the center of the valve and $s = \pm 1$ to it being on the posteromedial/anterolateral commissure of the valve respectively.% The short axes of the openings on each side of the clip were determined by keeping the maximum length of the leaflet edges equal to the length of the annulus line corresponding to each leaflet, and scaling the short axes of each opening in accordance with the relative scaling of the long axis of that opening.

\begin{table}[h!]
    \centering
    \caption{Position scale constants for the MVC simulations and resulting MV opening areas at E-wave peak. The MVC is positioned at $s\cdot l$ from the center of the coaptation line, with $l$ being the length of the long axis in the MV model. Here $s<0$ corresponds to clip positions anterolaterally of the center, and $s>0$ posteromedially. For reference, the area of the MV opening at mid diastole with no MVC is 2.54 cm$^2$.}
    \begin{tabular}{|c|c|c|c|c|c|}
        \hline
        $s$ & -0.4 & -0.2 & 0.0 & 0.2 & 0.4 \\
        \hline
        Area (cm$^2$) & 1.08 & 0.93 & 0.86 & 0.93 & 1.08 \\
        \hline
    \end{tabular}
    \label{table:clip_pos}
\end{table}

In addition to the five cases with the MVC in different positions, one case with an untreated valve was also simulated using the original MV model. Note that the untreated valve was modeled without any regurgitation. This is a model simplification motivated by the fact that our focus is to compare the blood flow into the ventricle during diastole. In this model the inflow is not greatly affected by any regurgitation in systole. The significant difference is that between the flow in the different double orifice cases in MVC treated valves and the single orifice of the untreated valve.

The LV used in these simulations has a volume at end diastole of $117$ ml, and an ejection fraction of 64~\%. These values, as well as the deformation of the ventricle volume throughout the heartbeats, are the same in all simulations. The untreated MV has a long axis of 12.8 mm, and short axes of 11.6 mm (posterior) and 8.1 mm (anterior) respectively. This results in an opening area of 4.0 cm$^2$ at E-wave peak, and 2.54 cm$^2$ at mid diastole. This is a relatively small opening area, so the resulting velocities and pressures in the MVC simulations are likely higher than a typical \textit{in-vivo} case.

All simulations were run on the KTH supercomputer \cite{beskow}, a Cray XC40 system  with 67,456 cores distributed over 4,120 Intel Xeon E5-269X 2.1/2.3 GHz CPUs. The simulations were run for six heartbeats, starting from early systole, all with identical wall motion from the captured TTE images. The time step length was set to 1 ms. One of the 26 surface meshes generated from the TTE data diverged significantly from the others, and was thus considered an outlier and excluded from the simulations. Between the rest of the captured frames, the surface meshes were temporally interpolated using cubic Hermite interpolation, to yield one surface mesh for each simulated time step.

\subsection{Blood flow analysis}

To analyze the blood flow in the ventricle we can visualize the flow velocity and pressure computed from the Navier-Stokes equations. 
It is also useful to analyze the velocity gradient $\nabla \mathbf{u}$, 
a tensor which describes the local spatial change in velocity at each time and position, defined by its components
\begin{equation}
    (\nabla \mathbf{u})_{ij} = \frac{\partial u_i}{\partial x_j}. 
\end{equation}
Near each point $\mathbf{x}_0\in \Omega^t$, the velocity gradient can be used to construct a linear approximation 
\begin{equation}
    \mathbf{u}(\mathbf{x}) \approx \mathbf{u}(\mathbf{x}_0) + \nabla \mathbf{u}(\mathbf{x}_0)(\mathbf{x}-\mathbf{x}_0), 
\end{equation}
and the mechanical stresses in the blood can be derived from the velocity gradient. 

A standard double decomposition of the velocity gradient constitutes a separation of the flow into a straining flow and a rotational flow, corresponding to its symmetric part $S(\mathbf{u})$, the strain rate tensor, and the skew-symmetric part $\Omega (\mathbf{u})$, the spin tensor:
\begin{equation}
    \nabla \mathbf{u} = \frac{1}{2} (\nabla \mathbf{u} + (\nabla \mathbf{u})^T ) + \frac{1}{2}(\nabla \mathbf{u} - (\nabla \mathbf{u})^T ) = S(\mathbf{u}) + \Omega (\mathbf{u}), 
\end{equation}
where the superscript $T$ represents the transpose of the tensor. The weakness of the double decomposition is that shear flow is unaccounted for. To address this shortcoming, the triple decomposition was proposed by \cite{kolavr2007vortex}, where the velocity gradient tensor is decomposed into a sum of three parts corresponding to irrotational straining flow (compression and elongation) $(\nabla \mathbf{u})_{EL}$, rigid body rotational flow $(\nabla \mathbf{u})_{RR}$, and shear flow $(\nabla \mathbf{u})_{SH}$:
\begin{equation}
    \nabla \mathbf{u} = Q ((\nabla \mathbf{u})_{EL} + (\nabla \mathbf{u})_{RR} + (\nabla \mathbf{u})_{SH})Q^T, 
\end{equation}
where $Q$ is an orthogonal matrix that represents a new frame of reference in which the shear flow can be subtracted from the other two parts. For the sake of brevity, we will refer to the three flow structures as strain, rotation and shear. 

Algebraically, the matrix decomposition may be interpreted as a sum of a normal symmetric part, a normal skew-symmetric part, and a non-normal part, which can be computed through the real Schur decomposition of the velocity gradient tensor, see e.g., \cite{van1996matrix}. 
In this article, the triple decomposition was computed by using the Matlab function \textit{schur}, which returns a frame of reference $Q$ that corresponds to a standardized form of the real Schur decomposition from which the shear flow tensor can be extracted, see \cite{hoffman2021energy}. 
To determine the magnitude of each flow component, Frobenius norms of each respective matrix were computed.

\section{Results}
\label{section:results}

%\todo[inline]{JK: Write a short overview of what results we present}

\subsection{Flow structures}

\begin{figure}[h!]
\begin{center}
\includegraphics[width=\linewidth]{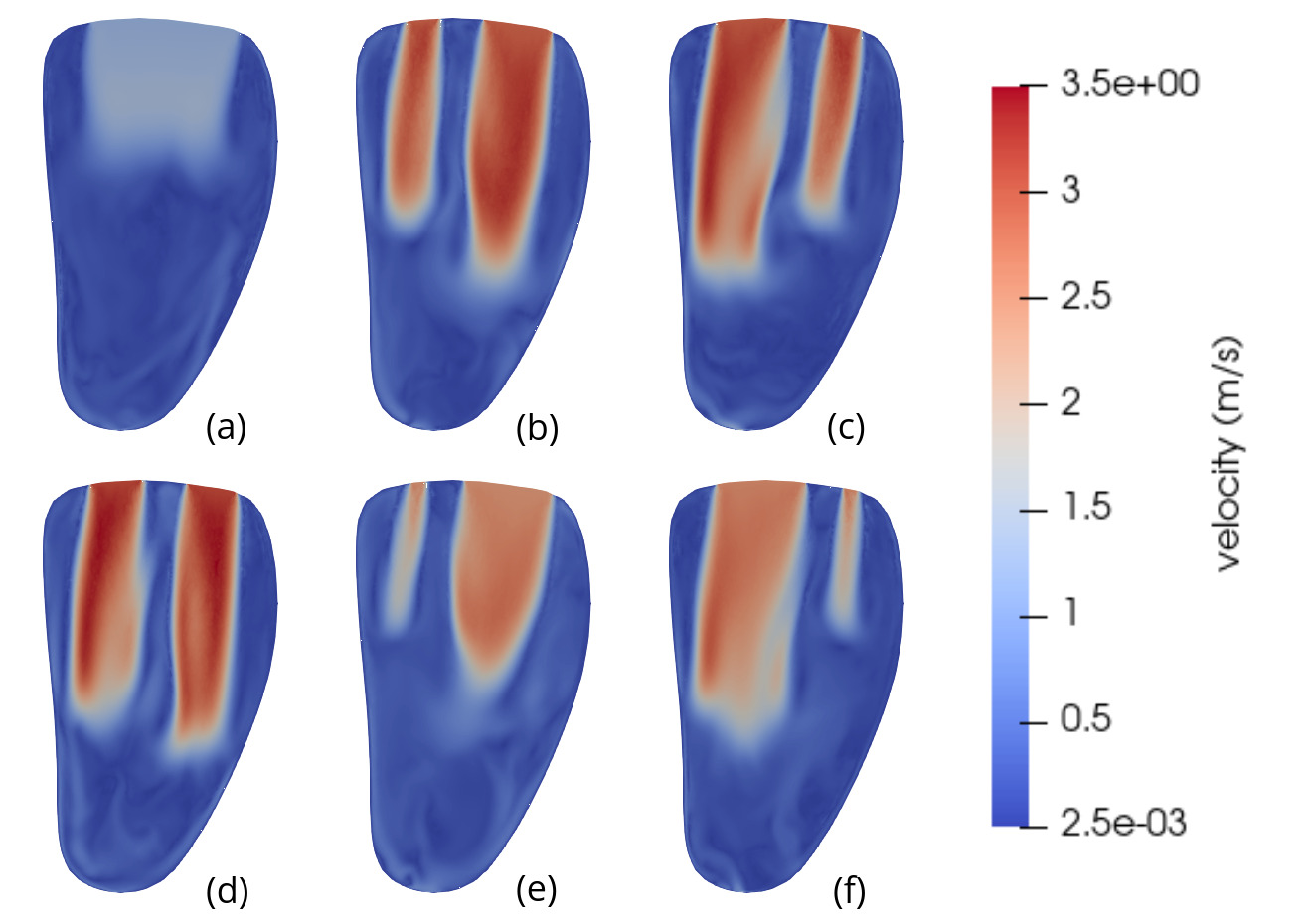}
\end{center}
\caption{Snapshots of the velocity magnitude at time $t = 0.45$ s in a vertical bicommissural cut through the left ventricle. This is in early diastole before the E-wave peak, when the intraventricular blood flow is dominated by strong inflow jets. Subfigure (a) shows the untreated case. The other subfigures have MVC's positioned at $s\cdot l$ from the center, with $l$ being the long axis of the MV model. The scale factor $s$ for the different cases are (b) $s=-0.2$, (c) $s=0.2$, (d) $s=0.0$, (e) $s=-0.4$, (f) $s=0.4$. The corresponding MV areas for each case in mid diastole are given in table \ref{table:clip_pos}.}
\label{fig:velocity}
\end{figure}

Snapshots of the velocity magnitudes in a bicommissural vertical cut through the ventricle for all simulated cases are shown in figure \ref{fig:velocity}. The snapshots are taken at time $t=0.45$ s from the start of the analyzed heartbeat (cf. figure \ref{fig:mvtimings}), i.e. 30 ms after the start of diastole. At this point in time inflow jets through the mitral valve can be observed in all cases. The untreated valve has a single wide jet, whereas the cases with MVC's in different positions show the double jets characteristic of inflow after MVC treatment.

The average velocity in the jet of the untreated case (figure \ref{fig:velocity}a), which has a larger MV opening area than the others, is smaller than the velocities in the clipped cases. This single wide jet also does not reach as far down into the ventricle as the jets in the other cases. Comparing the treated cases, the one with the clip in the center of the coaptation line (figure \ref{fig:velocity}d) has a slightly higher peak velocity than the others. This case is the one with the smallest total opening area, as given in table \ref{table:clip_pos}. In the ventricles with asymmetrically placed clips, the wider of the two jets typically reaches further into the ventricle than the smaller jet.

\begin{figure}[h!]
\begin{center}
\includegraphics[width=\linewidth]{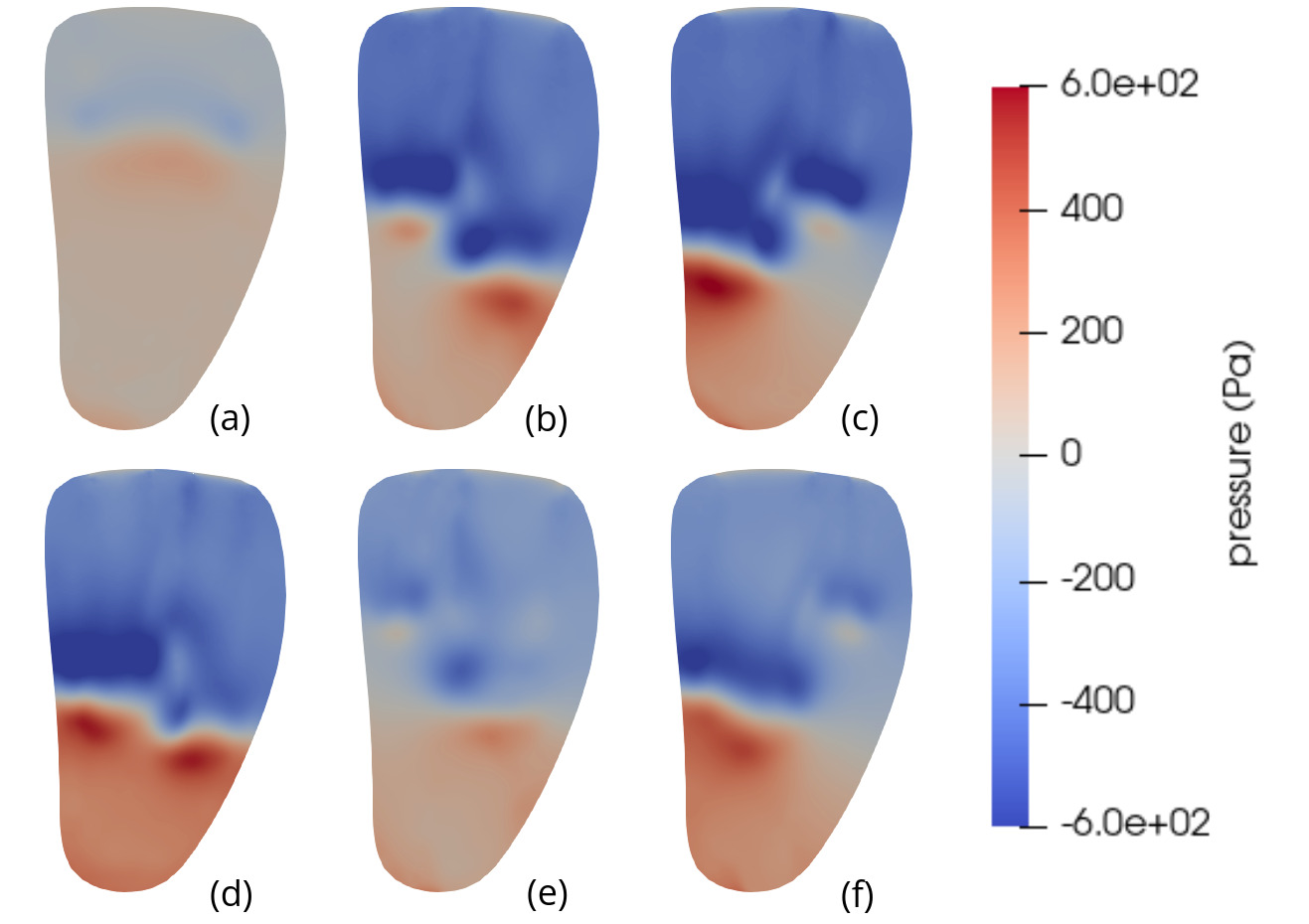}
\end{center}
\caption{Snapshots of the pressure at time $t = 0.45$ s in a vertical bicommissural cut through the left ventricle. This is in early diastole before the E-wave peak, when the intraventricular blood flow is dominated by strong inflow jets. The regions just below the jets experience high pressure, while there is a pressure drop around the base of the jet. (a) shows the untreated case. The other subfigures have MVC's positioned at $s\cdot l$ from the center, with $l$ being the long axis of the MV model. The scale factor $s$ for the different cases are (b) $s=-0.2$, (c) $s=0.2$, (d) $s=0.0$, (e) $s=-0.4$, (f) $s=0.4$. The corresponding MV areas for each case in mid diastole are given in table \ref{table:clip_pos}.} % Jets with higher velocity result in more pronounced pressure gradient.
\label{fig:pressure}
\end{figure}

Snapshots of the pressure at the same instant and the same cut as the velocity described above are shown in figure \ref{fig:pressure}. 
A recurring pattern of high pressure in the apical and mid-cavity parts of the ventricle is apparent in all simulated cases, with slight variations. In the untreated case (figure \ref{fig:pressure}a), where the jet does not extend as far into the ventricle, the overall pressure variations are lower than in the cases with MVC's. In the MVC cases, the double jets extend to different depths in the ventricle, which can also be seen in the velocity in figure \ref{fig:velocity}. This makes the line between high and low pressure less straight in the MVC cases, especially when the MVC is asymmetrically positioned.
High pressures just below the jets are apparent, but the lowest pressures are found around the front of the jets. This indicates the presence of vortex rings in these parts, since pressure is typically low in vortex centers.

%figure: pressure for all cases, similar to the velocity plot. Note low pressure in vortex centers and high pressure in high strain areas

\subsection{Triple decomposition of velocity gradient tensor}

Placing a clip in the MV significantly reduces the opening area of the MV, as can be seen in table \ref{table:results}. A natural result of this is higher inflow velocities, which the table also displays, with higher velocities the smaller the opening is. Both the velocity and the pressure are mainly affected by the reduction in area, but not so much by which side the clip is placed on. This is however not necessarily the case for the triple decomposition components, as is evident from the same table.

With the clip placed in the center of the MV, the opening area is the smallest out of all these simulation cases. For shear and strain, this leads to the highest computed peak values, but the peak rotation magnitude is actually slightly larger with the clip placed slightly posteromedially of the center. It is true for all three components that the peak values with the clip displaced to either side are not only dependent on how far the clip is displaced, but the direction as well. However, in most cases it does seem like smaller opening area does contribute to higher magnitudes of rotation, shear and strain, although the effect varies. The MV with the clip in the center has an opening area of 34~\% of that of the untreated MV. Its peak average rotation and strain are both roughly 190~\% of those for the untreated case, but its peak average shear is about 250~\% of that for the untreated case. As a reference, the maximum velocity increased about twofold in the same case. Throughout all simulation cases, shear has the highest relative increase when treated with MVC, whereas rotation and strain increase roughly equally.

\begin{table}[h!]
    \centering
    \caption{Peak values in the untreated simulation case, and the corresponding values for each of the MVC cases. Area is given at mid diastole. Velocity and pressure are the maximum values reached anywhere in the LV at E-wave peak. Rotation, shear and strain are the maxima reached in figure \ref{fig:integrated}.}
    \begin{tabular}{|l|c|c c c c c|}
        \hline
        Simulation case & untreated & -0.4 & -0.2 & 0.0 & 0.2 & 0.4 \\
        \hline
        Area at mid diastole (cm$^2$) & 2.54 & 1.08 & 0.93 & 0.86 & 0.93 & 1.08 \\
        \hline
        E-wave peak velocity (m/s) & 1.86 & 2.84 & 3.53 & 3.76 & 3.51 & 2.84 \\
        E-wave peak pressure (kPa) & 0.667 & 2.396 & 3.631 & 4.640 & 3.621 & 2.993 \\
        Peak average rotation (s$^{-1}$) & 21 & 35.4 & 37.0 & 39.8 & 39.9 & 34.1 \\
        Peak average shear (s$^{-1}$) & 174.6 & 344.8 & 397.5 & 439.1 & 409.7 & 368.7 \\
        Peak average strain (s$^{-1}$) & 51.4 & 82.5 & 86.7 & 95.3 & 93.6 & 81.3 \\
        \hline
    \end{tabular}
    \label{table:results}
\end{table}

\begin{figure}[h!]
\begin{center}
\includegraphics[width=10cm]{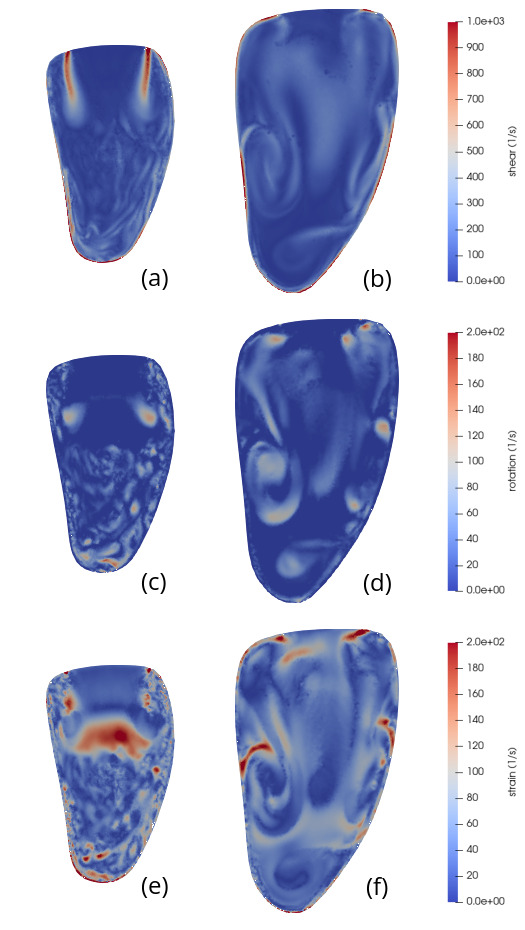}
\end{center}
\caption{Snapshots of shear flow (a and b), rotational flow (c and d) and straining flow (e and f) at times $t = 0.45$ s (a, c and e) and $t=0.71$ s (b, d and f) in a vertical bicommissural cut through the left ventricle. This simulation was done with an untreated valve. At time $t = 0.45$ s (early diastole), the flow is dominated by the incoming E-wave jet. The velocity and pressure at this time are shown in figures \ref{fig:velocity}a and \ref{fig:pressure}a respectively. At time $t = 0.71$ s (end diastasis) the inflow jet is weaker, and the initial dominant flow structures of the E-wave have started to break down. Note that the different sizes of the ventricle are due to the diastolic expansion.}%Next to the jet, high shear is experienced, while high strain appears just below the jet. At the front of the jet, vortex are formed, as seen in (c). At time $t = 0.71$ s there is not a clear flow pattern. Shear is focused on the wall, while some strain patterns arise from the wall to the interior of the ventricle. Rotation patterns along the ventricle are a reminiscent of the previous jet. Note that the different sizes of the ventricle are due to the diastolic expansion.}
\label{fig:triple_healthy}
\end{figure}

There are significant differences in how the flow in the ventricle behaves when dominated by strong inflow jets, i.e. during E-wave, and when there is no single dominating structure dictating most of the flow. Figure \ref{fig:triple_healthy} shows snapshots of the three modalities of the triple decomposition, i.e. rotation, shear and strain, in the simulation case without the MVC. The components are displayed at two different times in diastole: $t=0.45$ s, which corresponds to the velocities shown in figure \ref{fig:velocity}, and $t=0.71$ s, which is the time of the end of diastasis. Thus the first snapshots (figure \ref{fig:triple_healthy}a, c and e) show the triple decomposition at a point in time when the flow is dominated by the strong inflow through the MV in early diastole, and the second snapshots capture the components after the initial E-wave jet has broken down into smaller flow structures in the ventricle.

At $t=45$, it is clear that different flow modalities are prevalent in different parts of the ventricle. Rotation (figure \ref{fig:triple_healthy}c) appears in small connected areas in the lower part of the ventricle, but notably also on either side of the lower part of the jet structure, showing the presence of a vortex ring forming around the jet. Within the jet itself, where the flow is roughly unidirectional, the rotation is 0.

Shear (figure \ref{fig:triple_healthy}a) at the same moment is high at the outer edge of the jet, and along the lower wall of the ventricle. There are weaker shear structures throughout much of the rest of the ventricle as well, but none within the jet. Strain (figure \ref{fig:triple_healthy}e) is also large along the wall around the apical part of the ventricle. Additionally, below the jet, at the front of the blood flowing into the ventricle from the atrium, straining flow is high. There are also smaller spots throughout the ventricle with relatively high or low strain.
        
At $t=71$, when the E-wave jet has diminished and there is no single strong feature dominating the flow field, the overall distribution of the different flow modalities is more mixed. Rotation (figure \ref{fig:triple_healthy}d) is loosely connected in structures throughout the ventricle. Shear (figure \ref{fig:triple_healthy}b) and strain (figure \ref{fig:triple_healthy}f) are still relatively large along the ventricle wall around the apex, but the shear now also extends to other parts of the wall. In some structures near the wall but extending into the interior of the ventricle the strain is relatively high.

% Observations:
% D: clearly separated in space
% D: 0 rotation often corresponds to high strain, and v.v., except within the jet.
% D: No slip condition
% Comparison jet/no jet: magnitudes, locations, structures

\begin{figure}[H]
\begin{center}
\includegraphics[width=10cm]{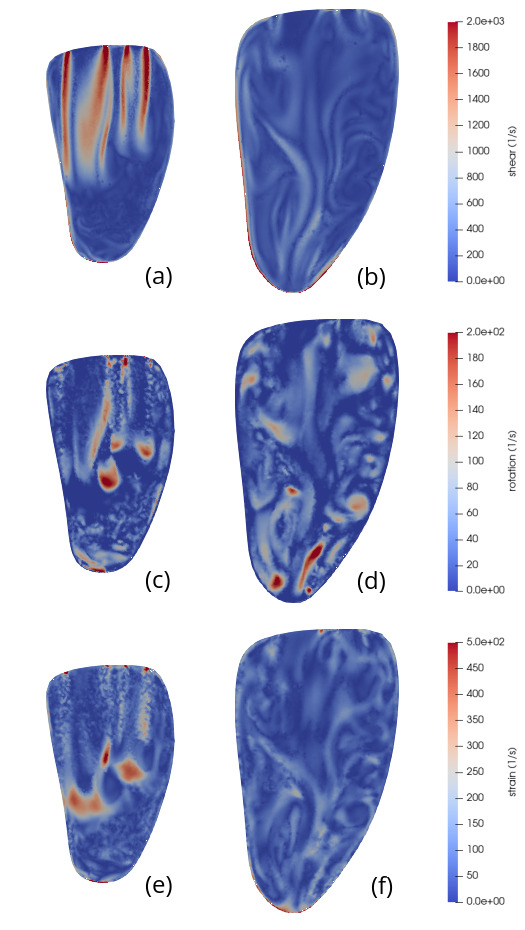}
\end{center}
\caption{Snapshots of shear (a and b), rotation (c and d) and strain (e and f) at times $t = 0.45$ s (a, c and e) and $t=0.71$ s (b, d and f) in a vertical bicommissural cut through the left ventricle. This simulation was done with MVC placed at $s\cdot l$ from the center of the coaptation line, with $l$ being the long axis in the MV model, and the position scale factor $s=0.2$. At time $t = 0.45$ s (early diastole), the flow is dominated by the two incoming E-wave jets. The velocity and pressure at this time are shown in figure \ref{fig:velocity}c and \ref{fig:pressure}c respectively. At time $t = 0.71$ s (end diastasis) the inflow jets are weaker, and the initial dominant flow structures of the E-wave have started to break down. Note that the different sizes of the ventricle are due to the diastolic expansion.}%Shear is experienced next to the jets, with peak values being twice as high as for the untreated valve. Again, high strain appears just below the jet while rotating structures appear at the front of the jets. Case (e) shows that the MVC case experiences higher strain that in an untreated valve. At time $t = 0.71$ s, shear is located next to the ventricle's wall and rotation and strain appear along the ventricle without forming clear patterns.}
\label{fig:triple_clip}
\end{figure}

When it comes to the cases treated with MVC's, some of the appearing flow structures share similarities with those in the untreated case. Snapshots of rotation, shear and strain for the MVC case with position scaling $s=0.2$ (corresponding to case c in figures \ref{fig:velocity} and \ref{fig:pressure}) are displayed in figure \ref{fig:triple_clip}. The snapshots are taken at the same moments in time as the ones in figure \ref{fig:triple_healthy}, and show the same cut through the ventricle. Note however the different scales for shear and strain.

With the flow profile dominated by two jets, as in early diastole in figure \ref{fig:triple_clip}a, c and e, the rotation (c) again shows structures reminiscent of vortex rings around the lower part of the jets. (For the corresponding velocity snapshot, see figure \ref{fig:velocity}c.) However, the magnitude of the vortex rings appear larger for the parts in the interior of the ventricle than where the jet runs close to the wall.

Along the edges of the jets, high shear is visible in the clipped case as well (figure \ref{fig:triple_clip}a). The shear of the wider jet seems to extend into the center of the jet as well, but this is because the displayed cut is not perfectly through the center of the jet. Like in the untreated case, there is also significant shear along the wall in the lower part of the ventricle.

The strain in the clipped case (figure \ref{fig:triple_clip}e) is once again large at the bottom of the jet where the inflow meets slower moving blood. It is also high at the wall in the apical part of the ventricle, and at the edges of the MV openings. There is also an area of higher strain right in the center of the ventricle, between the two jets.

At the time of the second snapshot, i.e. the end of diastasis when the strongest effects of the initial E-wave jets have diminished, the rotation structures are fairly well distributed in the ventricle, with some areas in the apical part of the ventricle where the rotation is relatively high (figure \ref{fig:triple_clip}d). Much like in the untreated case, the shear (figure \ref{fig:triple_clip}b) extends along more parts of the ventricle wall in the second snapshot, with weaker effects in the interior part. The strain (figure \ref{fig:triple_clip}f) retains its high magnitude along the apical part of the ventricular wall.

\begin{figure}[H]
\begin{center}
\includegraphics[width=15cm]{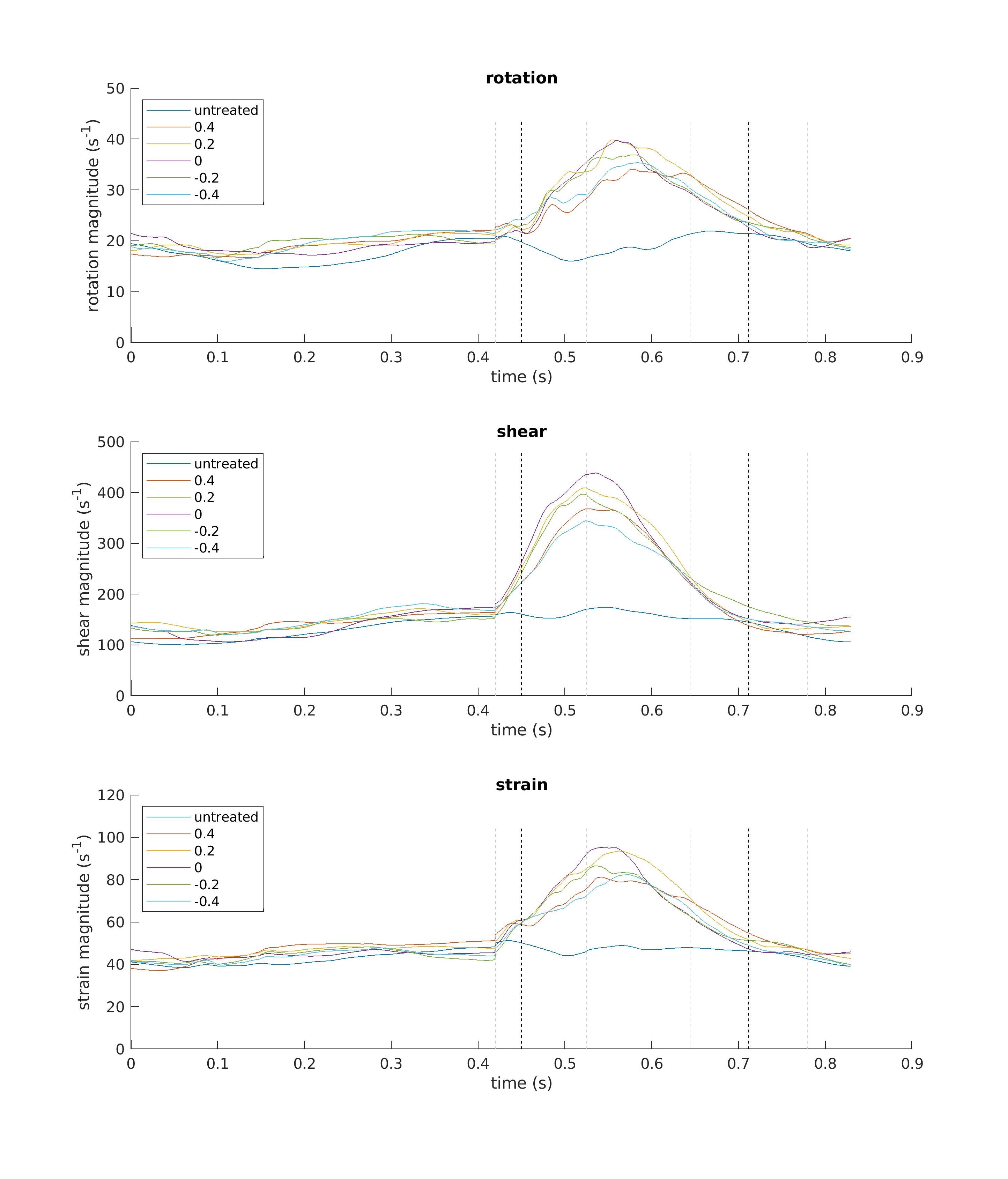}
\end{center}
\caption{Magnitudes of rotation, shear and strain averaged over the entire ventricle volume and plotted against time over one heartbeat, starting with systole, for all simulated cases. The legend indicates the location of the MVC as relative shift from the center of the coaptation line, except for the healthy (untreated) case. The gray dashed lines mark the time of the start of diastole, E-wave peak, diastasis start, and A-wave peak, also given in table \ref{table:diastole_timings}. The black dashed lines mark $t=0.45$ s and $t=0.71$ s, i.e. the times at which the blood flow is visualized in figures \ref{fig:velocity}, \ref{fig:pressure}, \ref{fig:triple_healthy}, \ref{fig:triple_clip}, \ref{fig:3D} and \ref{fig:vorticity}.}
\label{fig:integrated}
\end{figure}

% Move this paragraph to discussion? I think so
Differences between the triple decomposition components in the treated and untreated cases include, perhaps most notably, significantly higher magnitudes in strain and shear in the clipped case. The magnitude scales here have been adjusted to better show the location and structures of the different modalities, but figure \ref{fig:integrated} indicates that the claim holds. In the snapshots of the early diastolic flow, interactions between the two jets in the clipped case give rise to effects not seen in the untreated case, such as the high strain in the center of the ventricle. But vortex rings, high shear along the edge of the jets, and high strain at the bottom of the jets appear in both the treated and untreated cases. Shear and strain along the walls are also prevalent in both cases.

%Ideal place for fig 7_clip_schur_text

To get an idea of the duration and intensity of the different flow modalities from the triple decomposition over a full heartbeat, figure \ref{fig:integrated} shows the magnitude of the rotation, shear and strain for all simulated cases over time. There are some similarities apparent between all modalities. In systole, they all vary much less than in diastole, with the exception of the untreated case. Across all modalities, the magnitudes in the untreated case peak at much lower values than the MVC cases, despite being at similar levels in systole. It is difficult to distinguish any clear differences between the simulation cases in systole.

In diastole, all modalities follow similar temporal patterns in all MVC cases. With the increasing influx of blood in early diastole, the magnitudes of the rotation, shearand strain rise. Shortly after the E-wave peak, the magnitudes also peak and then start to fall. They continue decreasing towards systolic levels, with no real noticeable effects from the A-wave. However, despite having the same inflow volume of blood as the MitraClip cases, the magnitudes of rotation, shear and strain do not rise high above the systolic levels in the untreated case. In fact, after a small initial rise right at the start of diastole, they drop during some of the time leading up to the E-wave peak.

% Ideal place for fig 8_triple_integrated

\section{Discussion}
\label{section:discussion}

%\todo[inline]{Different flow modalities are highly localized. Strain and shear may induce thrombosis, and their effects are different => triple decomposition is a good tool. Shear is large along the wall, but is no-slip accurate?
%Future work: stream/streak/path lines to quantify effects on individual cells. Add atrium to examine pressure gradient.}

The triple decomposition used here gives an excellent view of the locality and structure of different flow modalities in the LV blood flow. Structures such as strong inflow jets can be visualized and quantified by studying the velocity, and with the addition of pressure, phenomena like the vortex rings around jets can be visualized. However, studying e.g. the shear around the jets requires more sophisticated tools. The triple decomposition captures this, as well as confirms expectations from the velocity and pressure plots. The pressure in the untreated case is relatively low in the basal part of the ventricle, but slightly higher further down. This pattern repeats in the MVC cases, and some of these have even lower pressure where the vortex ring around the inflow jet is expected. The rotation parts (a and b) of figures \ref{fig:triple_healthy} and \ref{fig:triple_clip} confirm that the rotation magnitude is indeed large in those parts. The high strain in the apical part of the ventricle also corresponds well to the high pressure in the same area. A related interesting feature of the rotation and strain is that areas with high rotation often tend to correspond to low strain, and vice versa. An exception to this observation is within the jets, where rotation is 0 and strain is also relatively low. See e.g. figure \ref{fig:triple_healthy}c and~e.

\begin{figure}[h!]
\begin{center}
\includegraphics[width=12cm]{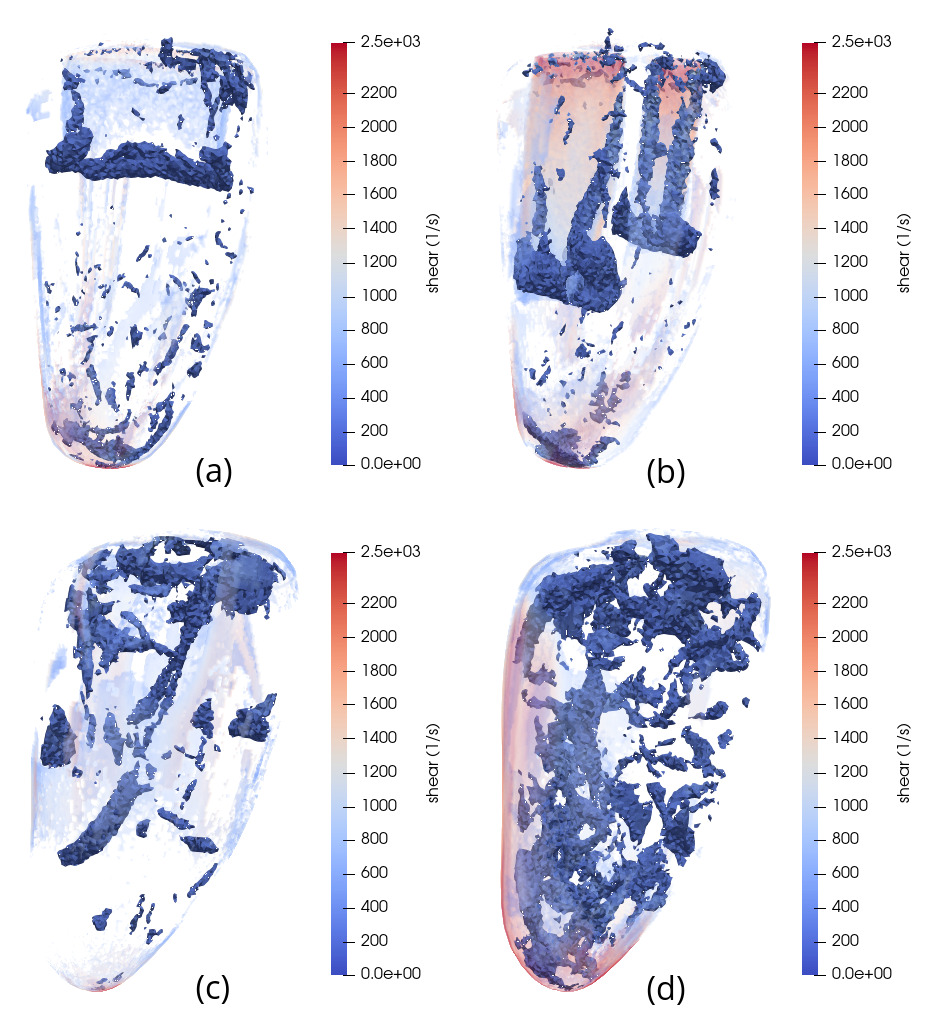}
\end{center}
\caption{Snapshots of 3D flow structures in the LV blood flow, as determined by the rotation magnitude. The opaque blue structures are all cells that have a rotation magnitude greater than 80 s$^{-1}$. The transparent structures are cells with shear magnitude greater than 500 s$^{-1}$, colored in accordance with the color bars. (a) Untreated case, $t=0.45$ s. (b) MVC case, $s=0.2$, $t=0.45$ s. (c) Untreated case, $t=0.71$ s. (d) MVC case, $s=0.2$, $t=0.71$ s. Corresponding 2D snapshots are displayed in figures \ref{fig:triple_healthy} and \ref{fig:triple_clip}.}
\label{fig:3D}
\end{figure}

To get a better idea of some of the 3D structures that appear in the LV flow, and how they develop over time, figure \ref{fig:3D} displays snapshots of some 3D structures defined by the computed rotation. Both the untreated and MVC case display clear vortex rings forming around the fronts of the jets as connected structures with rotation magnitude greater than a threshold value of 80 s$^{-1}$ (figure \ref{fig:3D}a and b). Above these the shear structures are visible at the border around the jet, with significantly higher magnitude in the MVC case. In the rest of the ventricle at this point in time both simulation cases display relatively little rotation, with only small connected volumes reaching above the threshold.

% TODO: This time is wrong! It should be t=0.71. Change during revision
At $t=0.71$ s, when the E-wave jets have decomposed, figure \ref{fig:3D}c and d gives a very qualitative view of the distribution of high rotation and shear structures throughout the ventricle. In the untreated case (c), the rotation appears as connected features, but with very little in the apical part of the ventricle. The shear is most prominent along the wall. In the clipped case (d), the rotation structures appear more numerous, and throughout the entire ventricle in different sizes. The shear is again concentrated on the wall, with higher values as indicated by the color bar. Interestingly, figure \ref{fig:integrated} at this point in time indicates that the total amount of rotation/shear magnitudes are similar in both the untreated and clipped ($s=0.2$) case. We have not investigated this further, but the reason could be that the more intense patterns visible in the MVC case make up a relatively small part of the entire ventricle volume, and that the integrated values in figure \ref{fig:integrated} are thus more strongly influenced by large regions of smaller values.

From figure \ref{fig:3D} the differences between the distributions of high rotation and shear structures at different points in time is evident. During the E-wave,  much is centered around the jets. Later in diastole, there is a wider distribution over the ventricle.

Since high shear is known to contribute to thrombosis formation, being able to quantify this in the interventricular blood flow is desirable. Our results suggest that the locations of high shear structures in diastole are similar before and after MVC treatment. They appear mainly around the inflow jets and along the ventricle walls. However, after MVC treatment the magnitude of the shear is higher, possibly as a result of the higher inflow velocities caused by the smaller MV opening areas. As seen in figure \ref{fig:integrated}, the integrated magnitude over the ventricle is higher for all MVC cases than for the untreated case throughout most of the E-wave and diastasis. This implies that the risk of platelets experiencing higher shear which could lead to thrombosis formation is increased after MVC treatment, but further studies would be required to quantify this risk.

While shear has been clearly linked to thrombosis formation, the case is more open for strain. Our results indicate that the locations of strain structures, like shear, do not change much in the LV after MVC treatment. But as for shear, strain is higher after treatment, both in maximum magnitude and integrated over the ventricle during the E-wave and diastasis.

Since shear, strain and rotation affect platelets differently, it is useful to separate these modalities using the triple decomposition, and be able to study them separately. Often vorticity has been used to similar ends. However, vorticity fails to separate rotation from shear, which can lead to misleading conclusions. Rotation only transports the blood cells, and thus likely does not contribute to platelet activation. Our results clearly show that both shear and rotation are present in the simulated blood flow, but in very different locations and of different magnitudes. Rotation is roughly one order of magnitude lower than shear, and is more prevalent in the vortex ring around the jet, and in connected vortex structures throughout the interior of the ventricle after the strong E-wave jet has diminished.

\begin{figure}[h!]
\begin{center}
\includegraphics[width=12cm]{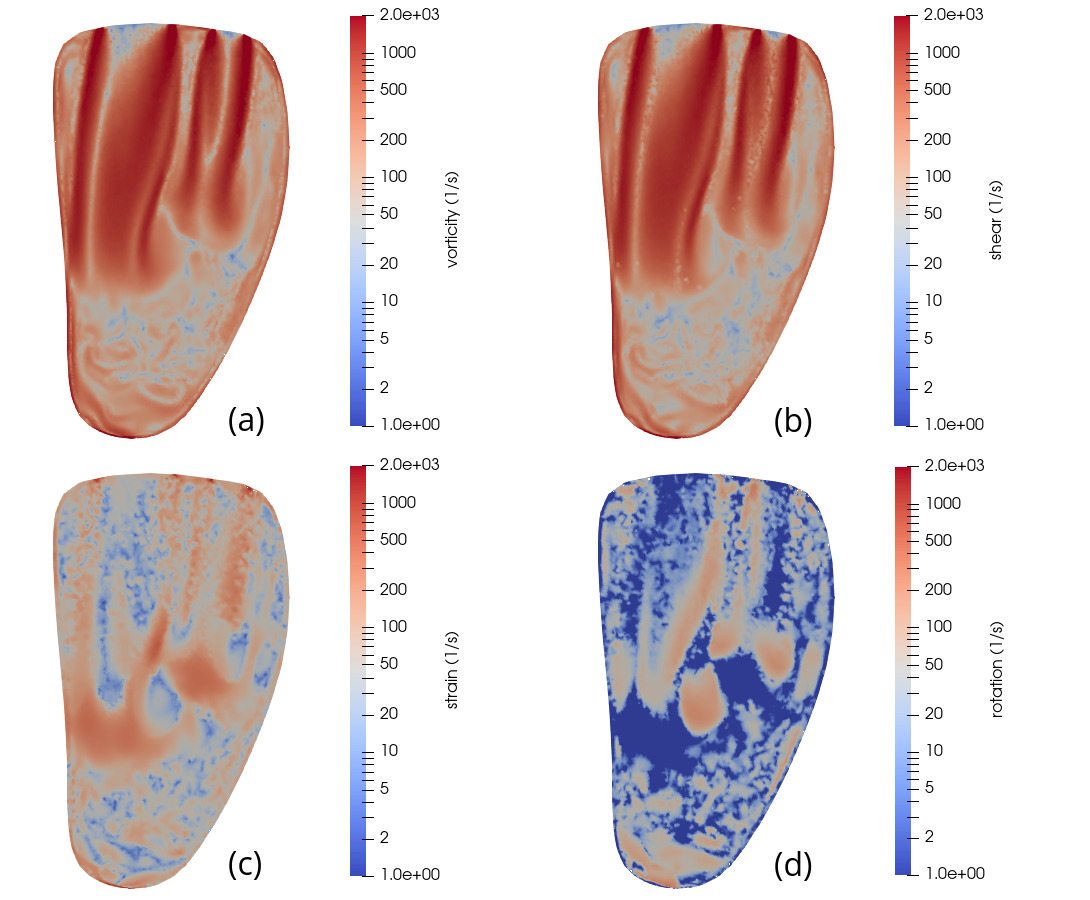}
\end{center}
\caption{Snapshots of (a) vorticity, (b) shear, (c) strain and (d) rotation at time $t = 0.45$ s in a bicommissural vertical cut through the left ventricle. This simulation was done with a MVC placed at $s\cdot l$ from the center of the coaptation line, with $l$ being the long axis in the MV model, and the position scale factor $s=0.2$. The velocity and pressure at this time are shown in figure \ref{fig:velocity}c and \ref{fig:pressure}c respectively.}
\label{fig:vorticity}
\end{figure}

To make a visual comparison between the vorticity and the triple decomposition, figure \ref{fig:vorticity} shows a snapshot of the vorticity for one of the simulated MVC cases, along with the three components from the triple decomposition at the same instant (same as in figure \ref{fig:triple_clip}a, c and e, but with a logarithmic scale). From this figure it is clear that the vorticity shares similarities with the shear. They reach similar magnitudes, and form structures mainly around the inflow jets and along the walls of the ventricle near the apex. But the logarithmic scale allows for observing the rotation contribution to the flow in the vorticity as well. E.g. the vortex rings show quite clearly as connected areas of high rotation magnitude around the bottom of the jets in figure \ref{fig:vorticity}d. (Note also the low strain in the vortex rings in figure \ref{fig:vorticity}c.) Since the vortex rings are rotation dominated, they do not show up as clear structures in the shear part (figure \ref{fig:vorticity}b). They do however clearly influence the vorticity magnitude around the lower parts of the jets in figure \ref{fig:vorticity}a, especially the large rotation structure in the center of the ventricle. Thus using vorticity as a measure of mechanical stimuli on blood cells may lead to inaccurate conclusions. In this particular case the magnitude of the shear is large in comparison to the rotation, which might limit the errors if vorticity were to be used in studies as a substitute for shear. In other settings this may not be the case, or one might be more interested in studying the rotation without influence of the shear. Either way, the triple decomposition offers a very promising way to distinguish rotation and shear in a way that vorticity fails to do.

% Observations:
% Vorticity mixes different modalities
% Here the magnitude is different, suppressing some modalities
% Effects on blood cells difficult to determine from vorticity

The simulated velocity in the untreated case (figure \ref{fig:velocity}) is within a reasonable range of a healthy heart, but the peak velocities in the clipped cases may be indicative of mitral stenosis \cite{aha_guidelines2021}. The opening areas of the clipped cases (table \ref{table:clip_pos}) are also quite small, and MVC treatment would likely not be clinically considered to mitigate regurgitation if the resulting areas were so small. Nevertheless, increases in inflow velocity as well as the triple decomposition components in the LV following MVC treatment are still likely to appear in clinically realistic cases due to decreased MV area.

While our results show that MVC treatment increases the magnitudes of rotation, shear and strain within the LV in early diastole, more information is needed to make any claims about the risk of platelet activation or thrombosis as a consequence of MVC treatment. Not only the magnitude of the shear and strain are needed to assess such risks, but the duration for which platelets experience these higher levels are of interest as well. However, studying the magnitude and locations of different flow component structures are an important first step, which opens up for new opportunities to learn more about mechanical stimuli in blood flow.

\section*{Conflict of Interest Statement}
%All financial, commercial or other relationships that might be perceived by the academic community as representing a potential conflict of interest must be disclosed. If no such relationship exists, authors will be asked to confirm the following statement: 

The authors declare that the research was conducted in the absence of any commercial or financial relationships that could be construed as a potential conflict of interest.

\section*{Author Contributions}
J.K. and J.H. designed the study and the triple decomposition algorithm, and wrote the article. J.K. ran the simulations and implemented the triple decomposition algorithm. F.S. designed and implemented the MV model, and D.L. designed and created the diastole timing and axis scaling tool for the MV model. S.E.L. and L.L. designed and implemented the MVC adaptation of the MV model, and designed the simulation cases. C.H.P., J.K. and J.H. post-processed and analyzed the results. C.H.P. made the figures in the results section. J.H. is the PI of the project.

%The Author Contributions section is mandatory for all articles, including articles by sole authors. If an appropriate statement is not provided on submission, a standard one will be inserted during the production process. The Author Contributions statement must describe the contributions of individual authors referred to by their initials and, in doing so, all authors agree to be accountable for the content of the work. Please see  \href{http://home.frontiersin.org/about/author-guidelines#AuthorandContributors}{here} for full authorship criteria.

\section*{Funding and acknowledgements}
The authors acknowledge the financial support from the Swedish Research Council (grant number 2018-04854), and the access to computing infrastructure by the Swedish National Infrastructure for Computing (grant number SNIC 2018/3-296).

Dr. Lucor has been funded in part by the TOR mobility program of the French Institute of Sweden, dedicated to the improvement of the scientific cooperation between France and Sweden.

%\section*{Acknowledgments}
%This is a short text to acknowledge the contributions of specific colleagues, institutions, or agencies that aided the efforts of the authors.

%\section*{Supplemental Data}
% \href{http://home.frontiersin.org/about/author-guidelines#SupplementaryMaterial}{Supplementary Material} should be uploaded separately on submission, if there are Supplementary Figures, please include the caption in the same file as the figure. LaTeX Supplementary Material templates can be found in the Frontiers LaTeX folder.

%\section*{Data Availability Statement}
%The datasets [GENERATED/ANALYZED] for this study can be found in the [NAME OF REPOSITORY] [LINK].
% Please see the availability of data guidelines for more information, at https://www.frontiersin.org/about/author-guidelines#AvailabilityofData

\bibliographystyle{frontiersinSCNS_ENG_HUMS} % for Science, Engineering and Humanities and Social Sciences articles, for Humanities and Social Sciences articles please include page numbers in the in-text citations
\bibliography{test}

\end{document}